\documentclass[12pt]{article}
\usepackage{amsmath,amssymb,epsfig}
\makeatletter \@addtoreset{equation}{section} \makeatother
\renewcommand{\theequation}{\thesection.\arabic{equation}}
\newcommand{\ba}{\begin{array}}
\newcommand{\ea}{\end{array}}
\newcommand{\beq}{\begin{equation}}
\newcommand{\eeq}{\end{equation}}
\newcommand{\bea}{\begin{eqnarray}}
\newcommand{\eea}{\end{eqnarray}}

\def\bce{\begin{center}}
\def\ece{\end{center}}

\def\nonu{\nonumber}

\def\pa{\partial}

\def\be{\beta}

\def\R{{\bf R}}

\def\S{{\bf S}}

\def\eps6{{\displaystyle \mathop{\epsilon}^{6}}{}}

\def\nab6{{\displaystyle \mathop{\nabla}^{6}}{}}

\def\0{{\sst{(0)}}}
\def\1{{\sst{(1)}}}
\def\2{{\sst{(2)}}}
\def\3{{\sst{(3)}}}
\def\4{{\sst{(4)}}}
\def\5{{\sst{(5)}}}
\def\6{{\sst{(6)}}}
\def\7{{\sst{(7)}}}
\def\8{{\sst{(8)}}}

\def\ba{\begin{array}}
\def\ea{\end{array}}
\def\beq{\begin{equation}}
\def\eeq{\end{equation}}
\def\be{\begin{equation}}
\def\ee{\end{equation}}

\def\eps{\epsilon}

\def\ba{\begin{array}}
\def\ea{\end{array}}
\def\beq{\begin{equation}}
\def\eeq{\end{equation}}
\def\be{\begin{equation}}
\def\ee{\end{equation}}

\def\eps{\epsilon}

\newcommand{\bean}{\begin{eqnarray*}}
\newcommand{\eean}{\end{eqnarray*}}

\begin{document}
\thispagestyle{empty} \addtocounter{page}{-1}
\begin{flushright}
{\tt hep-th/0606073}\\
\end{flushright}

\vspace*{1.3cm}

\centerline{ \Large \bf Two Circular Wilson Loops and Marginal Deformations}
\vspace{.3cm} 
\centerline{ \Large \bf  } 
\vspace*{1.5cm}
\centerline{{\bf Changhyun Ahn}} 
\vspace*{1.0cm} 
\centerline{\it
Department of Physics, Kyungpook National University, Taegu
702-701, Korea} 
\vspace*{0.8cm} 
\centerline{\tt ahn@knu.ac.kr} 
\vskip2cm

\centerline{\bf Abstract}
\vspace*{0.5cm}

We study type IIB supergravity backgrounds which are dual to
marginal deformations of ${\cal N}=4$ super Yang-Mills theory. 
We re-examine two circular Wilson loops and describe how the phase 
transition occurs in the presence of deformation parameter. 

\baselineskip=18pt
\newpage
\renewcommand{\theequation}
{\arabic{section}\mbox{.}\arabic{equation}}

\section{Introduction}

The solution-generating technique \cite{LM} provides a new gravity 
solution which is dual to marginally deformed field theories.
The deformed solution preserves ${\cal N}=1$ supersymmetry as long as
the direction corresponding to $U(1)_R$ R-symmetry is not involved
in this procedure. 

This method can be also used to find the gravity dual of 
deformed Coulomb branch RG flow of ${\cal N}=4$ super Yang-Mills theory
where this part of moduli space corresponds to a continuous 
distribution of D3-branes on an ellipsoidal shell
\cite{AV}.
The UV limit of the dual gauge theory is the Leigh-Strassler deformation
\cite{LS} of ${\cal N}=4$ super Yang-Mills theory. 

Recently, 
in \cite{AV1}, for certain moduli space, the $\sigma$ deformation induces
a transition from Coulombic attraction between quark and anti-quark to
linear confinement where the scale of confinement increases with this
deformation parameter $\sigma$.
Moreover, this method can also be used to the case of massive 
quark and monopole \cite{Ahn} and the Wilson loop computation  
implies that either this deformation enhances the Coulombic attraction or
it induces a phase transition to linear confinement.

Gross and Ooguri \cite{GO} have found a phase transition
where the classical minimal surface can have a topology of annulus
or consists of two disconnected surfaces. When the distance between the loops
are very small, then the area of annulus is smaller than the one of
disconnected surfaces. As the distance increases, the area of annulus also
increases. At some critical distance, 
the disconnected surface becomes more dorminant.   
This jump from one saddle point to the other should lead to 
a phase transition 
in the Wilson loop correlator.
This configuration was studied in $AdS_5$ space in \cite{Zarembo99}
by solving the equations of motion 
where two concentric circles of equal radii were considered. 
For certain value of distance, the classical connected solution ceases to
exist. In other words, the connected minimal surface becomes unstable at this
value. The exact critical point where the areas of connected surface 
and disconnected surface are equal to each other 
is less than this value. For the unequal radii, similar analysis was done 
in \cite{OZ} and the finite temperature case was analyzed in \cite{KPTM}.

In this paper, we re-compute two circular Wilson loop case originated from 
\cite{Zarembo99} in the marginally
deformed $AdS_5 \times \S^5$ type IIB background.
After describing equal radii, then we also study the different radii case.
Starting with Lunin-Maldacena deformed metric \cite{LM}, one can construct 
Nambu-Goto action and its equations of motion.  Given the appropriate
boundary conditions, the distance between the two loops is a function
of an integration constant by elliptic integrals and we present its
behavior
under the deformation parameter explicitly. 
Similarly, the area of minimal connected and disconnected surfaces 
can be constructed and we
describe its behavior with respect to the distance between the loops
by changing the
deformation
parameter.      

First, 
we have found the ``deformed'' expression for the distance $L$ between 
two circular Wilson loops in terms of the radius $R$ and an
integration constant $k$ by (\ref{LF}), (\ref{fka1}) and (\ref{para}).
The expression for $L$ is basically the same as the one in \cite{Zarembo99}
with modified integration constant. 
Secondly, we also have found the ``deformed'' energy of minimal
surfaces
for connected and disconnected ones by (\ref{S}) and (\ref{disc})
respectively.
Contrary to the $L$, 
the expression for $E$ cannot be obtained by simply modifying 
an integration constant due to an extra overall  factor with 
deformation parameter. This is kind of new observation. 
They can be described by two figures, Figure 1 and Figure 2.
We have extended our result to the unequal radii case and they are
given by (\ref{h}), (\ref{fka1}) 
and (\ref{action}) together with Figure 3 and
Figure 4. 
Finally, we computed the on-shell action of D5-brane characterized by
(\ref{a}).

\section{Two circular Wilson loops revisited}

The Wilson loop 
correlator can be expressed as an area of the classical string worldsheet
stretched between the loops.
Let us consider two Wilson loops in the boundary of $AdS_5$ 
and they are concentric 
circles of radius $R$(or $R_1$ and $R_2$ for unequal radii) 
separated by a distance $L$ \cite{GO,Zarembo99}.

The Lunin-Maldacena deformation \cite{LM} 
of $AdS_5 \times S^5$ background of 
type IIB theory has the string frame metric 
\bea
ds_{str}^2 = \alpha' \sqrt{H} \left( U^2 d x_{\mu}^2 +
 \frac{dU^2}{U^2} +  ds_{\tilde{S}^5}^2 \right), \qquad 
H = 1 + \hat \sigma^2, \qquad \hat\sigma \equiv \sigma/2
\nonu
\eea 
where we set the length scale of $AdS_5$ to one.
The conformal factor $H\geq 1$ 
becomes a constant \cite{AV1} after using the 
equations of motion for internal coordinates on five-sphere
$\tilde{S}^5$
which depends on the modulus of complex $\beta$. This $\beta$
can be realized by two real deformation 
parameters $\gamma$ and $\sigma$. 
The string tension $1/(2\pi \alpha')$ is proportional to
the Yang-Mills coupling $g_{YM}$.
The Euclidean 
$AdS_5$ metric can be written in terms of cylindrical coordinates 
\cite{Zarembo99} which are appropriate for the symmetry of
Wilson loops above we are considering, 
by using a change of variable  $z \equiv 1/U$(the $AdS_5$ 
boundary is located at $z=0$),
\bea
ds_E^2 = \alpha' \frac{\sqrt{H}}{z^2} \left( dt^2 + dz^2 + dx^2 + dr^2 +
r^2 d \phi^2 \right).
\nonu
\eea
The Nambu-Goto action for a fundamental string on the type IIB 
supergravity 
background from the ansatz \cite{Zarembo99} 
for the minimal surface $t=0, \phi=\sigma, 
r=r(\tau), x=x(\tau)$ and $z=z(\tau)$ is 
\bea
S = 2\pi \int d \tau \frac{r}{z^2} \sqrt{H} \sqrt{x'^2 + r'^2 + z'^2}
\label{nbaction}
\eea
where $'$ denotes a derivative with respect to $\tau$.
Note that the $\hat\sigma=0$ limit reproduces the undeformed result
\cite{Zarembo99} because $H=1$. 

We would like to study the effects of $\hat\sigma$ deformation on the two
Wilson loops by following the procedure of \cite{Zarembo99}.
Since the action does not explicitly depend on $x$, the equation of motion
for $x$ implies 
\bea
\frac{r}{z^2} \frac{x'}{\sqrt{x'^2 + r'^2 + z'^2}} \sqrt{H} \equiv k
\label{constant}
\eea
which is an integration constant. Note the presence of 
a factor of $\sqrt{H}$ which will propagate all the remaining computations. 
For the positive $k$, $x'$ is also positive
and one can choose the gauge $\tau=x$ \footnote{For $k=0$, the minimal surface
in $AdS_5$ bounded by a circle of radius $R$ is $r^2+z^2 =R^2$ 
\cite{DGO1,BCFM}. We will see this case from the discussion of
(\ref{S}) by taking $k=0$ limit. }.
Then the Euler-Lagrange equations of motion for $x,r$ and $z$ 
with this gauge choice 
can be summarized as
\bea
r'^2 + z'^2 + 1  - \frac{r^2}{k^2 z^4} H & = & 0, \nonu \\ 
r'' - \frac{r}{k^2 z^4} H & = & 0, \nonu \\
z'' + \frac{2 r^2}{k^2 z^5} H & = & 0
\label{equations}
\eea
where  the constant of equation of motion (\ref{constant}) 
is used in the second and third equations.
Compared with the undeformed theory \cite{Zarembo99}, 
the dependence on the deformation
parameter $H$ appears in these equations.
Whenever we need to know the undeformed results, we simply put 
$H$ as one.

\subsection{The loops have equal radii}

We assume that two circular Wilson loops are located at 
$x=\pm L/2$ on the $AdS_5$ boundary located at $z=0$. 
Then the boundary conditions
for the differential equations (\ref{equations}) 
are characterized by 
\cite{Zarembo99}
\bea
r(-L/2) = r(L/2) = R, \qquad 
z(-L/2) = z(L/2) = 0.
\nonu
\eea
$R$ and $L$ are the radii of the circular Wilson loops and 
the distance between them respectively.
The modified boundary condition of $z$ will be discussed later.
As done in \cite{Zarembo99} explicitly, 
the solutions for the last two equations of (\ref{equations}) 
satisfying the boundary conditions 
are given by 
\bea
r = \sqrt{a^2-x^2} \cos \theta, \qquad
z = \sqrt{a^2-x^2} \sin \theta, \qquad a^2 \equiv R^2 +\frac{L^2}{4}.
\label{eq}
\eea
Here the parametric 
angle $\theta$ from the first equation of (\ref{equations}) 
satisfies
\bea
\theta' = \pm \frac{a}{a^2-x^2} \sqrt{\frac{H \cos^2 \theta}{k^2 a ^2 \sin^4
\theta}-1}
\label{thetaprime}
\eea
where the upper sign is for the negative $x$ and the lower sign is for
the positive $x$. From the above boundary conditions, it is easy to see
that $\theta(-L/2)=\theta(L/2)=0$ and note the presence of a factor $H$
inside of the square root. Strictly speaking, the modified 
boundary condition due to the regularization will be present and will be
used later. 

By using an integral formula \cite{GR,KPTM},
\bea
\int_{0}^{\theta} 
d \theta  \frac{\sin^2 \theta}{\sqrt{\cos^2 \theta -\frac{k^2 a^2 
\sin^4 \theta}{H}}} =
\frac{\sqrt{H}}{ka} \frac{\beta_{+}-1}{\sqrt{\beta_{+}-\beta_{-}}} 
\left[ \Pi\left(\chi,\frac{1-\beta_{-}}
{\beta_{+}-\beta_{-}},
\kappa \right)-F(\chi, \kappa)\right]
\label{thetaexpression}
\eea
where $\Pi$ and $F$ are the elliptic integrals of the third and first
kind, respectively and let us introduce various ``deformed'' parameters 
\cite{KPTM} in the sense that we are dealing with 
$H$-dependent quantities
\bea
\beta_{\pm} & = & \frac{\frac{2k^2 a^2}{H}+ 1 \pm \sqrt{1+\frac{4k^2 a^2}{H}}}
{\frac{2k^2 a^2}{H}}, \qquad
\chi = \sin^{-1} \sqrt{\frac{(\beta_{+}-\beta_{-})(1-\cos^2 \theta)}
{(1-\beta_{-})(\beta_{+}-\cos^2 \theta)}}, \nonu \\
\kappa & = & \sqrt{\frac{\beta_{+}(1-\beta_{-})}{\beta_{+}-\beta_{-}}},
\label{para}
\eea
the relation (\ref{thetaprime}) provides
that the expression (\ref{thetaexpression}) is equal to
\bea
\frac{\sqrt{H}}{2 k a} \ln \frac{(a+\frac{L}{2})(a \pm x)}{(a-\frac{L}{2})
(a \mp x)}
\label{function}
\eea
where 
the upper sign is for the region $ -\frac{L}{2} \leq x \leq 0$
and the lower sign is for the region $ 0 \leq x \leq \frac{L}{2}$.
From (\ref{function}), one can check that $\theta(-L/2)=\theta(L/2)=0$.
Note that in the computation of an integral (\ref{thetaexpression}),
it is crucial to identify the relative magnitudes of four roots of
the denominator and numerator of an integrand(where we made a change
of variable $\cos^2 \theta$ as new one in (\ref{thetaexpression})) 
and lower limit of an integral. 
In our case, although the presence of the deformation parameter 
$H$ appears in (\ref{para}) explicitly, 
it is easy to see the relative magnitudes of these five quantities.
As in undeformed case \cite{Zarembo99}, 
there exists a relation $\beta_{+} > 1 \geq 
\cos^2 \theta \geq \beta_{-} > 0$.

In particular, at $x=0$, 
by eliminating the common factor $\frac{\sqrt{H}}{k a}$, one arrives at
\bea
F(ka) = \frac{1}{2} \ln \left( \frac{a+\frac{L}{2}}{a-\frac{L}{2}}
\right) =
\ln \left( \frac{\sqrt{R^2+\frac{L^2}{4}}+\frac{L}{2}}{R}\right)
\label{fka}
\eea
where we substituted the expression of $a$ in (\ref{eq}),
the function $F(ka)$ is defined by (\ref{thetaexpression}) without 
a factor $\frac{\sqrt{H}}{2 k a}$ and this can be reduced to
``deformed'' function
\bea
F(ka) =\frac{\beta_{+}-1}{\sqrt{\beta_{+}-\beta_{-}}} 
\left[ \Pi\left(\frac{1-\beta_{-}}
{\beta_{+}-\beta_{-}},
\kappa \right)-K(\kappa)\right]
\label{fka1}
\eea
together with ``deformed'' parameters (\ref{para}).
Here we use the fact that the parametr $\chi$ in (\ref{para}) becomes 
$\frac{\pi}{2}$ when $\theta$ at $x=0$ satisfies $\cos^2 \theta = 
\beta_{-}$ \cite{Zarembo99}. 
This enables us to write $F(ka)$ in terms of 
the complete elliptic integrals as above. 
After differentiating this $F(ka)$ with respect to $ka$ and putting it
to
zero, one gets $K(\kappa)=2E(\kappa)$ 
where the explicit form of $\kappa^2$ in (\ref{para}) is
\bea
\kappa^2 =\frac{1}{2}\left( 1+
  \frac{1}{\sqrt{1+4\frac{k^2a^2}{1+\hat\sigma^2}}}
\right).
\label{kappa2}
\eea
Numerically, this $\kappa^2$ becomes $0.826$ satisfying the condition 
$K(\kappa)=2E(\kappa)$ and this leads to 
\bea
k a =0.58 \sqrt{H}
=0.58 \sqrt{1+\hat\sigma^2}. 
\label{specialka}
\eea 
The $\hat\sigma$ deformations increase $ka$ for fixed distance $L$ between 
the two circular Wilson loops. 
See Figure 1 for details.
For large $ka$, the above $F(ka)$ has the following 
asymptotic behavior, by realizing that the third kind of 
complete elliptic integral can be reduced to the second kind one,
\bea
F(ka) =  \frac{H^{1/4}}{\sqrt{2ka}} 
\left[ 2E(\frac{1}{\sqrt{2}})-K(\frac{1}{\sqrt{2}})\right]
=  \frac{H^{1/4}}{\sqrt{2ka}}  \frac{2\pi^{3/2}}{\Gamma^2(\frac{1}{4})}
\label{fkaka}
\eea 
which approaches zero as $ka$ becomes very large.
Due to the deformation parameter $H^{1/4}$, the undeformed 
case approaches to zero faster than deformed cases.
For small $ka$, one can expand the elliptic integrals around
$ka=0$ and arrives at $F(ka) =-\frac{ka}{\sqrt{1+\hat\sigma^2}} 
\ln \frac{ka}{\sqrt{1+\hat\sigma^2}}$. Therefore, the function $F(ka)$
becomes zero at $ka=0$ and due to a factor $1/\sqrt{H}$, the ``deformed''
$F(ka)$ approaches zero faster than undeformed $F(ka)$.
See also Figure 2.

Finally, by simplifying (\ref{fka}), one gets   
``deformed'' relation between the distance $L$ and an integration constant
$k$
\bea 
L = 2R \sinh F(ka)
\label{LF}
\eea
with ``deformed'' function $F(ka)$ (\ref{fka1}). 
Although the functional relation of (\ref{LF})
looks similar to the one for undeformed theory \cite{Zarembo99}, 
note that 
the dependence on the deformation 
parameter arises from $\beta_{\pm}$ and $\kappa$ in (\ref{para}).

Figure 1 shows the $ka$ dependence of $L$ when $R=1$.
For each undeformed and ``deformed'' case, 
there exists a maximal distance $L_{max}$ between 
the two circular Wilson loops. 
This $L_{max}$ can be obtained by substituting 
(\ref{specialka}) with (\ref{LF}) and is given by 
\bea
L_{max}=1.04R
\label{lmax}
\eea
where the corresponding integration constant is
$ka=0.58 \sqrt{1+\hat\sigma^2}$ and 
the two branches from large 
$ka$ and from small $ka$  meet at these
points. As we pointed out, at these two extreme cases
of $ka$ the distance $L$ approaches to zero.
Note that the deformation parameter $H=1+\hat\sigma^2$ is cancelled
out in (\ref{LF}) after plugging (\ref{specialka}) and the above
maximal distance $L_{max}$ is the same as the one in undeformed theory
\cite{Zarembo99}. This is clear from Figure 1. 
If $L > L_{max}$, then the classical connected 
solution becomes unstable \cite{GO}
and 
the physical solutions are two discontinous ones \cite{GO}. 
As we observed, when $ka=0$, the distance $L$ goes to zero.
In other words, a single circular Wilson loop case can be seen at $L=0$.
This will be discussed after the energy (\ref{S}) is determined later.
When $ka$ is very large, 
the $\hat\sigma$ deformation increases the value of $ka$ for 
fixed $L$ and furthermore increases $L$ for fixed $ka$.

Since the area of connected surface needs to be regularized \cite{Zarembo99}, 
the boundary
condition for $z$ can be modified as $z(\pm L/2)=\epsilon$. Accordingly, 
the boundary condition for $\theta(x)$ is changed to $\theta(\pm L/2)=
\tan^{-1} (\frac{\epsilon}{R})$ from (\ref{eq}) \cite{Zarembo99}.   
The area of the regularized connected 
surface can be written as \footnote{In this 
computation we use the following integral formula \cite{GR}:
\bea
\int^{U}_{C} \sqrt{\frac{X-D}{(A-X)(B-X)^3(X-C)}} dX & = &
\frac{2(C-D) F(\Delta, Q)}{(B-C)\sqrt{(A-C)(B-D)}} -\frac{2
\sqrt{(A-C)(B-D)} E(\Delta,Q)}{(A-B)(B-C)} \nonu \\
&+& \frac{2(B-D)}{(A-B)(B-C)} 
\sqrt{\frac{(A-U)(U-C)}{(B-U)(U-D)}}
\nonu 
\eea
where $A>B>U>C>D$, $\Delta=
\sin^{-1} \sqrt{\frac{(B-D)(U-C)}{(B-C)(U-D)}}$ and $Q=
\sqrt{\frac{(B-C)(A-D)}{(A-C)(B-D)}}$ where $F$ and $E$ are elliptic 
integrals of first and second kind. Then one can easily see the 
divergent part of $S$ which originates from the last term of the right hand 
side above. }
\bea
S &=& 2\pi \int_{-\frac{L}{2}}^{\frac{L}{2}}
 dx \frac{r}{z^2} \sqrt{H} \sqrt{1 + r'^2 + z'^2}
=4\pi \int_{\frac{\epsilon}{R}}^{\theta(x=0) } d \theta 
\frac{\sqrt{H} \cot^2 \theta}{\sqrt{\cos^2 \theta - 
\frac{k^2 a^2 \sin^4 \theta}{H}}}
\nonu \\
&=& 
4\pi \sqrt{H}\left( 1 + \frac{4k^2 a^2}{H}\right)^{1/4}
\left[ (1-\kappa^2) K(\kappa) - E(\kappa) \right]+
\frac{4\pi \sqrt{H} R}{\epsilon} 
\label{S}
\eea
where ``deformed'' $\kappa$ is the same as the one in (\ref{para}) or 
(\ref{kappa2}).
We used the equations of motion (\ref{constant}), (\ref{eq}),
and (\ref{thetaprime}) in this computation. 
Compared with the undeformed case \cite{Zarembo99}, 
the overall factor $\sqrt{H}$ appears and this will lead to
an overall shift of $S$. 
Moreover, the first two terms of (\ref{S}) have 
``deformed'' parameter and this will 
change the slope of Figure 2 significantly as the 
$\hat\sigma$ increases.
Therefore, two circular Wilson loops correlator 
can be obtained from (\ref{fka1}), (\ref{LF}), and (\ref{S}).
Some results from gauge theory side in perturbation theory 
were found in \cite{PS,APS}.

\begin{figure}[ht]
   \epsfxsize=4.5in \centerline{\epsffile{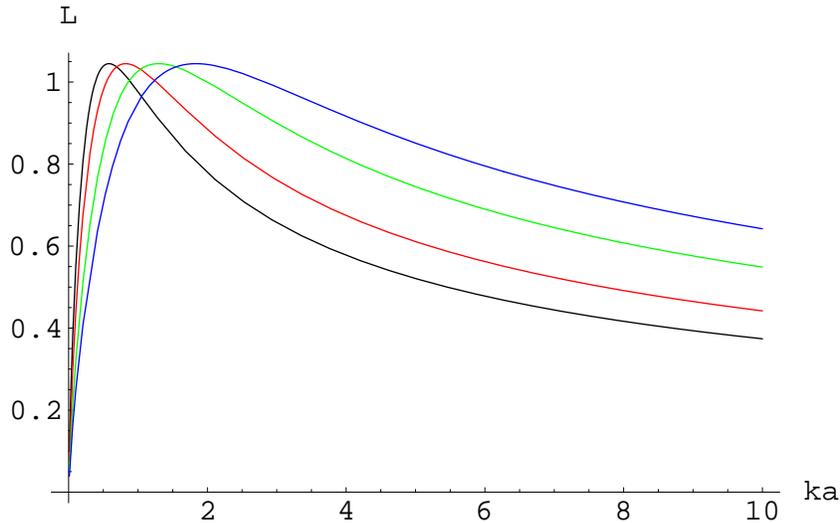}}
   \caption[FIG. \arabic{figure}.]{ The $L$ dependence of 
$ka$ using (\ref{LF}) with (\ref{fka1}) 
for $\hat\sigma =0$(black), 1(red), 2(green),
and
3(blue). These curves for large $ka$ appear from bottom to top.
The maximal distance $L_{max}$ is given by 
$1.04R$ (\ref{lmax}) 
where we set $R=1$ and the corresponding $ka$'s with
(\ref{specialka}) 
are 
$0.58$, $0.82=0.58\sqrt{2}$, $1.30=0.58\sqrt{5}$, $1.83=0.58\sqrt{10}$ 
respectively. 
When $ka$ is large, 
the $\hat\sigma$ deformation increases the value of $ka$ for 
fixed $L$ and increases $L$ for fixed $ka$. 
Note that the maximal
distance is the same both undeformed($\hat\sigma=0$) case and 
deformed($\hat\sigma \neq 0$) case. The whole curve  moves to the
right hand side as $\hat\sigma$ increases.}
\label{fig1}
\end{figure}

For large $ka$(in other words, $L$ goes to zero) 
where $\kappa $ goes to $\frac{1}{\sqrt{2}}$ from 
(\ref{kappa2}), 
the area can be approximated as
\bea
S= \frac{4\pi \sqrt{H} R}{\epsilon} -\frac{16\pi^4 \sqrt{H}}{\Gamma^4(
\frac{1}{4})} \frac{R}{L} =\sqrt{H} \left( 
 \frac{4\pi  R}{\epsilon} -\frac{16\pi^4 }{\Gamma^4(
\frac{1}{4})} \frac{R}{L}
\right).
\label{limitS}
\eea
Here we write $ka$ in terms of $L$ through (\ref{fkaka}) and (\ref{LF}).
This is exactly the minimal surface
for anti-parallel lines \cite{Malda,RY}, 
each of length $R$ and separated by a distance 
$L$ with ``deformed'' parameter \cite{AV1}.
This indicates that when we consider for the two circular Wilson loop case 
of ellipsoidal D3-brane distribution with deformation, 
the similar limiting procedure as above will lead to 
the minimal surface of deformation of Coulomb branch flow
\cite{AV1}. We will comment on this possibility next section.
Therefore, the area of minimal surface is increased by 
a factor $\sqrt{H}$ which is greater than or equal to 1.

When $ka=0$, $\kappa=1$ and $E(1)=1$. 
Then the regular part of (\ref{S}) becomes 
$-4\pi \sqrt{H}$ which will be the regular piece of disconnected surface 
below (\ref{disc}).

On the other hand,
the area of ``deformed'' regularized disconnected surface    
is obtained from the descrpition of \cite{DGO1,BCFM}
by multiplying a factor $\sqrt{H}$ because the metric has an extra
factor $\sqrt{H}$ which is a constant
\bea
S_{disc.} = -4\pi\sqrt{H} + \frac{4\pi \sqrt{H} R}{\epsilon}.
\label{disc}
\eea
When the regular piece of connected surface and the one for
disconnected surface are equal to each other, one can solve 
the condition for $ka$ numerically. By equating the first two terms of 
(\ref{S}) and the first term of (\ref{disc}), one 
gets 
\bea
ka=1.31\sqrt{H}
=1.31\sqrt{1+\hat\sigma^2}
\label{cricon}
\eea 
for nonzero $ka$. 
By inserting this value into (\ref{LF}), 
one finds the critical distance is given by   
\bea
L_{cri.}=0.91R
\nonu 
\eea 
which is exactly the same as the one in undeformed
case \cite{Zarembo99} because the deformation parameter is cancelled
after substituting (\ref{cricon}) into (\ref{LF}).
This critical distance $L_{cri.}$ 
is less than the maximal distance $L_{max}$ (\ref{lmax}). 

\begin{figure}[ht]
   \epsfxsize=4.5in \centerline{\epsffile{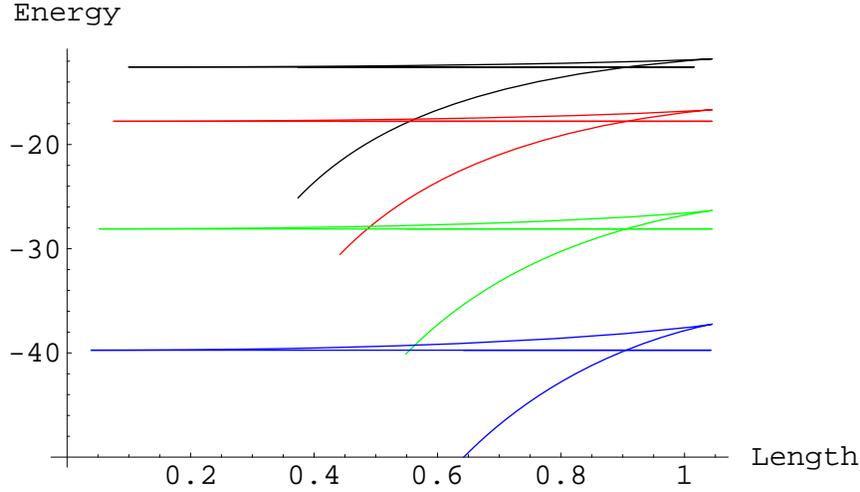}}
   \caption[FIG. \arabic{figure}.]{ The $L$ dependence of 
$S-\frac{4\pi \sqrt{H} R}{\epsilon} $ characterized by the curves 
and $-4\pi \sqrt{H}$ characterized by straight lines 
using (\ref{S}), (\ref{disc}), (\ref{LF}) and (\ref{fka1}) 
 for $\hat\sigma =0$(black), 1(red), 2(green),
and
3(blue)(from top to bottom). 
Note the appearance of cusps at the maximal distance $L_{max}=
1.04R$ where we set $R=1$. The connected surface intersects 
with disconnected surface at $L=0$ and $L_{cri.}=0.91R$. 
Recall that around $L=0$, as the deformation parameter increases,
$L$ approaches to zero faster(The minimum value of $L$ for blue one is less
than the minimum value $L$ of black one in the first branch while the 
opposite holds in the second branch).
The $\hat\sigma$ deformation enhances the energy of surface.   }
\label{fig2}
\end{figure}

Figure 2 describes the $L$ dependence of the regular parts of $S$ and 
$S_{disc.}$ showing the phase transition at the $L=L_{cri}$.
Using the parametric plot between these two quantities, we can eliminate 
the dependence of parameter $ka$ and obtain the dependence of
the regular parts of $S$ and $S_{disc.}$ on the distance $L$ directly.
We use the equations (\ref{S}), (\ref{disc}), (\ref{LF}) and (\ref{fka1}).  
There exist two kinds of branches. The first branch is located in the 
region between $L=0$ and $L_{max}$ where the cusp appears and the 
second branch is located in the region between $L_{max}$ and $L_{min}$
which is not equal to zero. These two branches for the regular piece
of $S$ intersect with the regular piece of $S_{disc.}$ at both 
$L_{cri.}=0.91R$ and $L=0$, as we explained
above.
The location of cusps(the difference between the energy 
for disconnected minimal surface and the one for connected surface at 
$L=L_{max}$) 
are increased as the deformation parameter
$\hat\sigma$ increases.

As the deformation parameter $\hat\sigma$ increases, the minimum value
$L_{min}$ of the first branch becomes smaller and 
the one of second branch becomes 
larger. This can be understood from the behavior of
$ka$ in the ``deformed'' function $F(ka)$, as we discussed previously:
a factor of $H^{-1/2}$ for small $ka$ and a factor of $H^{1/4}$ for 
large $ka$.  
Before the point $L_{cri.}=0.91R$ is reached, 
the classical 
connected solution has lower action than the disconnected one and
will dominate in the two circular Wilson loop correlator 
while after that critical point is reached, 
the disconnected solution will dominate. In other words, 
the Gross-Ooguri phase transition occurs across this critical point. 
For fixed length $L$, the slopes of the curve(the derivative 
an energy with respect to th distance) increase
as the $\hat\sigma$ increases. This fact reflects the area (\ref{S})
has ``deformed'' parameter. 
For the energies, the effect of $\hat\sigma$ enhances the 
strengths of energies
of minimal surface negatively.

\subsection{The loops have different radii}

Two circular Wilson loops are located at $x=0$ and $x=h$ on the 
$AdS_5$ boundary.
For the unequal radii, the boundary conditions  for 
equations (\ref{equations}) are \cite{OZ}
\bea
r(0)= R_2, \qquad r(h) = R_1, \qquad z(0) =z(h)=0.
\nonu
\eea
$R_1$ and $R_2$ are the radii of the two circular Wilson loops 
and $h$ is the distance between them.
The solutions with these boundary conditions 
can be written as \cite{OZ}
\bea
r & = & \sqrt{a^2-(x+c)^2} \cos \theta, \qquad
z = \sqrt{a^2-(x+c)^2} \sin \theta, \nonu \\ 
c & \equiv & \frac{R_2^2-R_1^2}{2h}-\frac{h}{2}, \qquad
a^2 \equiv c^2 +R_2^2
\nonu
\eea
where $a$ and $c$ are integration constants.
Here there exists a relation  
\bea
\theta' = 
\pm \frac{a}{a^2-(x+c)^2} \sqrt{\frac{H \cos^2 \theta}{k^2 a ^2 \sin^4
\theta}-1}
\nonu
\eea
where the upper sign is for $x$ in $0<x<x_0$ and lower sign 
is for $x$ in $x_0 < x< h$ for some $x_0$.
The $x_0$ and $h$ can be obtained from the following expressions after 
$x$-integrations: from $0$ to $x_0$ and from $x_0$ to $h$ respectively
\cite{OZ}
\bea
\frac{\sqrt{H}}{2 k a} \ln \frac{(a+x_0 +c)(a -c)}{(a-x_0-c)
(a + c)}, \qquad 
\frac{\sqrt{H}}{2 k a} \ln \frac{(a+h +c)(a -x_0-c)}{(a-h-c)
(a + x_0+c)}
\nonu
\eea
that are equal to $\theta$-integration (\ref{thetaexpression}) 
respectively.
By adding these one gets ``deformed'' function in terms of
$h,R_1$, and $R_2$
\bea
F(ka)  & = & \frac{1}{4} \ln \frac{(a+h +c)(a -c)}{(a-h-c)
(a + c)}  \nonu \\
&= & \frac{1}{2}
\ln \left( \frac{R_1^2+R_2^2+h^2 + \sqrt{(R_2^2-R_1^2)^2+
h^4+2h^2(R_1^2+R_2^2)}}{2R_1R_2} \right)
\label{rel}
\eea
where $F(ka)$ is the same as before (\ref{fka1}).

By simpifying this relation (\ref{rel}), 
the ``deformed'' distance of two circular Wilson loops for different 
radii corresponding to (\ref{LF}) for equal radii is given by 
\bea
h =R_2 \sqrt{2\alpha \left[1+2 \sinh^2 F(ka)\right]-\alpha^2 -1}
\label{h}
\eea
where the ratio of two radii is
\bea
\alpha \equiv R_1/R_2.
\nonu
\eea
As observed in \cite{OZ}, due to the positivity of
the inside of the square root, there exists some possible range for the 
ratio of two radii $\alpha$. The maximum value of $\alpha$ is
$2.72$ \cite{OZ}.
The maximum value of $h$ occurs when $\alpha= 1+ 2\sinh^2 F(ka)$.
Then $h_{max}$ can be obtained and it is given by 
\bea
h_{max}=1.17R_2
\label{hmax}
\eea 
under the 
condition (\ref{specialka}).
Of course, when $\alpha=1$(equal radii), the distance $h$ is reduced to 
(\ref{LF}) with $R=R_2$.
When both $h$ and $ka$ are equal to zero, the solution (\ref{h})
implies
that only $\alpha=1$ is valid. 
Other values of $\alpha$ cannot give 
simultaneous zero for $h$ and $ka$ because there exists an extra term
$(\alpha-1)^2$
inside of the square root of (\ref{h}) 
in addition to ``deformed'' function $F(ka)$
dependent term.   
As the deformation parameter $\hat\sigma$ increases, the minimum value
$h$ of the first branch becomes smaller and 
the one of second branch becomes 
larger.

One can draw the plot $ka$ versus $h$ and it turns out 
the behavior of the plot for fixed $\alpha$ looks similar  to Figure 1.
The main difference is the behavior of near $ka=0$. Of course this 
region has an unstable branch. Since we are 
considering $\alpha =1.5$ for unequal radii, contrary to the previous case,
for small $ka$, $h$ cannot be zero. 
We can assume that $\alpha$ is greater than 1 because for the region 
of $\alpha < 1$, 
there exists a symmetry under the inversion $R_1 \leftrightarrow R_2$
and one can use the result from the region of $\alpha > 1$.
As observed in Figure 3, the arc length 
from the lowest position near $ka=0$ to the highest position where the 
distance $h$ has its maximum value $h_{max}$ is less than the one in equal 
radii case. This can be seen from the next Figure 4 also.  
The behavior of $h$ near $ka=0$ and large $ka$ can be read off
from (\ref{h}) by inserting the asymptotic expression for $F(ka)$ at
these two regimes. Evidently, $h$ cannot be zero due to the presence
of $(\alpha-1)^2$ inside of the square root as we mentioned before.

\begin{figure}[ht]
   \epsfxsize=4.5in \centerline{\epsffile{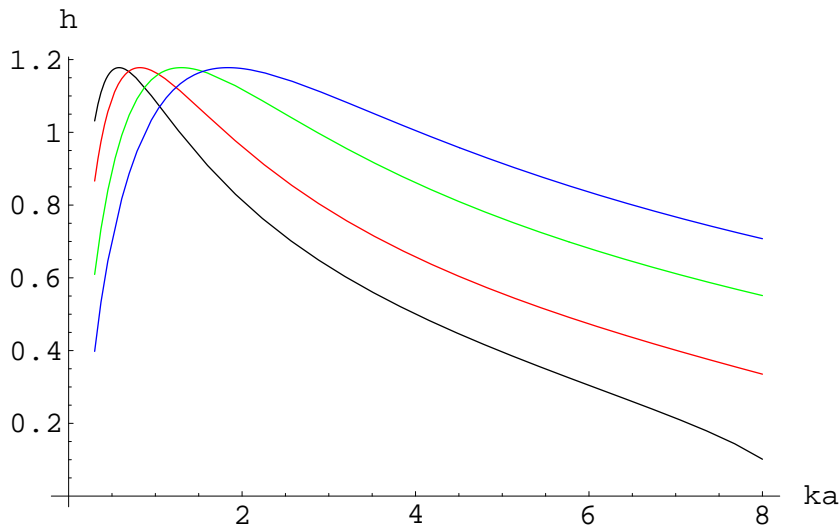}}
   \caption[FIG. \arabic{figure}.]{ The $h$ dependence of 
$ka$ with $\alpha=R_1/R_2=1.5$ 
using (\ref{h}) and (\ref{fka1}) for $\hat\sigma =0$(black), 1(red), 2(green),
and
3(blue). The maximal distance $h_{max}$ (\ref{hmax}) is given by 
$1.17R_2$ where we set $R_2=1$ and the corresponding $ka$'s with
(\ref{specialka}) 
are 
$0.58$, $0.82=0.58\sqrt{2}$, $1.30=0.58\sqrt{5}$, $1.83=0.58\sqrt{10}$ 
respectively. 
The $\hat\sigma$ deformation increases the value of $ka$ for fixed $h$. 
Note that the maximal
distance is the same both undeformed($\hat\sigma=0$) case and 
deformed($\hat\sigma \neq 0$) case. 
For $\alpha \neq 1$, the asymptotic behavior for large and small $ka$
leads to a nonzero $h$.
The whole curve  moves to the
right hand side as $\hat\sigma$ increases.}
\label{fig3}
\end{figure}

For the area of minimum surface, 
the boundary conditions for $\theta$ are characterized by
$\theta(x=0)=\tan^{-1} (\frac{\epsilon}{R_2})$ and $\theta(x=h)=
\tan^{-1} (\frac{\epsilon}{R_1})$ \cite{OZ}. 
The regularized area of the connected surface can be obtained similarly
\bea
S & = & 2\pi \int_{0}^{h}
 dx \frac{r}{z^2} \sqrt{H} \sqrt{1 + r'^2 + z'^2}
=2\pi \left( \int_{\frac{\epsilon}{R_2}}^{\theta(x=x_0)}
+ \int_{\frac{\epsilon}{R_1}}^{\theta(x=x_0)} \right) d \theta 
\frac{\sqrt{H} \cot^2 \theta}{\sqrt{\cos^2 \theta - 
\frac{k^2 a^2 \sin^4 \theta}{H}}}
\nonu \\
& = & 
4\pi \sqrt{H}\left( 1 + \frac{4k^2 a^2}{H}\right)^{1/4}
\left[ (1-\kappa^2) K(\kappa) - E(\kappa) \right]+
\frac{2\pi \sqrt{H} (R_1+R_2)}{\epsilon} 
\label{action}
\eea
where $\theta(x=x_0)=\cos^{-2} \beta_{-}$ with (\ref{para}) and 
the first two terms are the same expression for equal radii case.

For large $ka$, as we did previously, 
by rewriting $ka$ in terms of $\sinh F$(therefore $h, R_1$ and $R_2$
from (\ref{h}) in this case), the area can be written as
\bea
S= \frac{2\pi \sqrt{H} (R_1+R_2)}{\epsilon} -\frac{16\pi^4 \sqrt{H}}{\Gamma^4(
\frac{1}{4})} \sqrt{\frac{R_1R_2}{(R_1-R_2)^2+h^2}}
\nonu
\eea
which reduces to (\ref{limitS}) when $R_1=R_2$.
There is also an overall factor $\sqrt{H}$.
On the other hand, for small $ka$, as we observed before,
the regular part of (\ref{action}) becomes 
$-4\pi \sqrt{H}$ which is the regular piece of disconnected surface 
(\ref{disc}).

By equating the first two terms of 
(\ref{action}) and the first term of (\ref{disc}), one 
gets $ka=1.31\sqrt{1+\hat\sigma^2}$ for nonzero $ka$ which is the same
condition as equal radii case. 
By inserting this value into (\ref{h}), 
one finds 
\bea
h_{cri.}=0.99R_2
\nonu 
\eea 
which is exactly the same as the one in undeformed
case \cite{OZ} and 
is less than the maximal distance $h_{max}$ (\ref{hmax}). 

Figure 4 describes the $h$ dependence of the regular parts of $S$ and 
$S_{disc.}$ showing the phase transition at the $h=h_{cri.}$.
The regular piece
of $S$ intersects with the regular piece of $S_{disc.}$ at 
$h_{cri.}=0.99R_2$, as we explained
above.
As the deformation parameter $\hat\sigma$ increases, the minimum value
of the first branch becomes smaller and the one of second branch becomes 
larger, like as equal radii case. 
For the energies, the effect of $\hat\sigma$ enhances the 
strengths of energies of minimal surface.

\begin{figure}[ht]
   \epsfxsize=4.5in \centerline{\epsffile{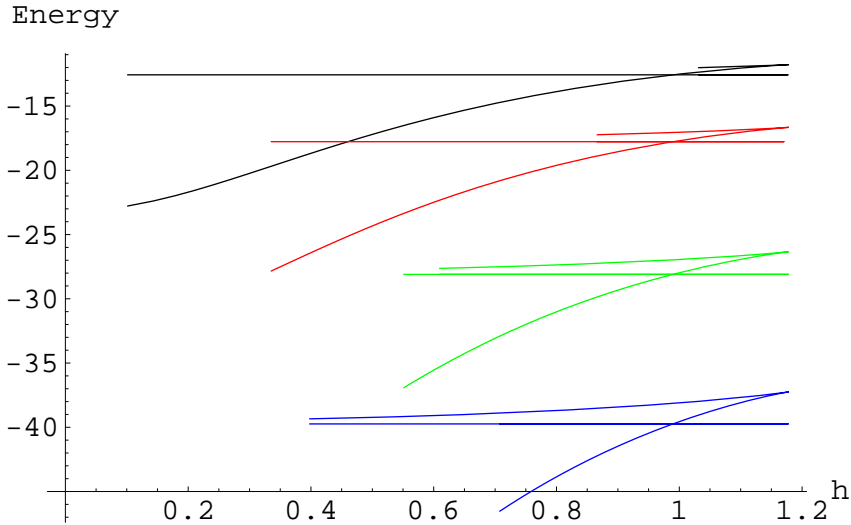}}
   \caption[FIG. \arabic{figure}.]{ The $h$ dependence of 
$S-\frac{2\pi \sqrt{H} (R_1+R_2)}{\epsilon} $ characterized by the curves 
and $-4\pi \sqrt{H}$ characterized by straight lines 
using (\ref{action}) and (\ref{disc}) 
for $\hat\sigma =0$(black), 1(red), 2(green),
and
3(blue). We take $\alpha=R_1/R_2=1.5$. 
Note the appearance of cusps at the maximal distance $h_{max}=
1.17R_2$ where we set $R_2=1$. The connected surface intersects 
with disconnected surface at $h_{cri.}=0.99 R_2$. 
Recall that around $h=0$, as the deformation parameter increases,
$h$ approaches to zero faster. 
The $\hat\sigma$ deformation enhances the area of surface.   }
\label{fig4}
\end{figure}

\section{On-shell action of D5-brane}

The Euclidean $AdS_5$ metric without Lunin-Maldacena 
deformation can be written as 
\bea
ds_{AdS_5}^2 = \frac{L^2}{y^2} \left( dy^2  + dr^2 +
r^2 d \sigma^2 +  dx^2 + dt^2 \right)
\nonu
\eea
which is appropriate for the symmetry of two circular Wilson loops as 
in previous section.
The $AdS_5$ boundary is locatd at $y=0$. The $L$ is the radius of 
$AdS_5$ and is given by $L^4 = \alpha'^2 4\pi g_s N$ where 
$g_s$ is a string coupling constant and $N$ is the number of D3-branes.
Note that we restored the radius of $AdS_5$ in this section 
and $y$ plays the
role of $z$ of previous section.

The action for the Euclidean probe D5-brane consists of 
Dirac-Born-Infeld part and Wess-Zumino part 
as follows
\bea
S_{bulk} = T_5 \int d^6 \xi \sqrt{\mbox{det} (G+ \cal {F} )} - i T_5 
\int {\cal F} \wedge C_4
\nonu
\eea
where $T_5$ is the tension of D5-brane and is given by
$T_5 = \frac{1}{(2\pi)^5 \alpha'^3 g_s}$ and the relevant part of 4-form
potential is $C_4 = L^4  \left(\frac{3}{2} \theta_k
-\sin 2\theta_k +\frac{1}{8} \sin 4\theta_k \right)$ multiplied by the volume
of unit 4-sphere $S^4$ that is $\frac{8}{3}\pi^2$ \cite{Yama}.
For the fundamental string charge of D5-brane, there should be an electric 
worldvolume gauge field. 

Here we take one of the spacetime coordinate as one of the 
worldvolume coordinate $\sigma$ and the other worldvolume coordinate 
is denoted by $\tau$. For the minimal surface we assume 
$t=0, r=r(\tau), x=x(\tau)$ and $y=y(\tau)$ as in previous section.
The D5-brane is wrapping on $S^4$ with $\theta=\theta_k$ which is a 
constant \cite{Yama}.
Evaluated on this ansatz,  
the action becomes
\bea
S_{bulk} = \int d \tau d \sigma  {\cal L}_{bulk}
\nonu
\eea
where
\bea
\frac{{\cal L}_{bulk}}{ T_5 \left( \frac{8}{3} \pi^2 \right) L^4 } =
\sin^4 \theta_k \sqrt{\frac{L^4 r^2 }{y^4} \left( r'^2 + x'^2 + y'^2 
\right) + {\cal F}_{\tau \sigma}^2 } - 
i {\cal F}_{\tau \sigma} \left(\frac{3}{2} \theta_k
-\sin 2\theta_k +\frac{1}{8} \sin 4\theta_k \right)
\nonu
\eea
where $'$ denotes a derivative with respect to $\tau$.
The conjugate momentum to the coordinate $y$ is 
\bea
p_y = \frac{\pa {\cal L}_{bulk}}{\pa y'} =
\frac{ T_5 \left( \frac{8}{3} \pi^2 \right) L^4 \sin^4 \theta_k 
\frac{L^4 r^2 }{y^4} y' }{\sqrt{\frac{L^4 r^2 }{y^4} 
\left( r'^2 + x'^2 + y'^2 
\right) + {\cal F}_{\tau \sigma}^2 }}
\label{py}
\eea
and the conjugate momentum to the gauge field $A_{\sigma}$ is conserved since
$A_{\sigma}$ does not appear in the action, is equal to the 
fundamental string charge of D5-brane with $-i$ and is 
given by
\bea
p_A & = & 2\pi\alpha' \frac{\pa {\cal L}_{bulk}}{\pa {\cal F}_{\tau
    \sigma}} 
\nonu
\\
&=&
  2\pi\alpha'  T_5  \frac{8}{3} \pi^2  L^4 
\left[ \frac{\sin^4 \theta_k {\cal F}_{\tau \sigma}}
{ \sqrt{\frac{L^4 r^2 }{y^4} \left( r'^2 + x'^2 + y'^2 
\right) + {\cal F}_{\tau \sigma}^2}} - i \left(\frac{3}{2} \theta_k
-\sin 2\theta_k +\frac{1}{8} \sin 4\theta_k \right) \right].
\label{pa}
\eea

Since the action does not depend on the coordinate $x$,
the equation of motion for $x$ implies 
\bea
\frac{ T_5  \frac{8}{3} \pi^2  L^4 \sin^4 \theta_k 
\frac{L^4 r^2 }{y^4}x' }{\sqrt{\frac{L^4 r^2 }{y^4} 
\left( r'^2 + x'^2 + y'^2 
\right) + {\cal F}_{\tau \sigma}^2}}  \equiv K
\label{constants}
\eea
and one can choose the gauge $\tau=x$.
The corresponding Euler-Lagrange equations of motion for 
$x, r$ and $y$ can be written as
\bea
r'^2 + y'^2 + 1  - \frac{\left( T_5  \frac{8}{3} \pi^2 
    L^4 
\sin^4 \theta_k  \right)^2 
L^4 r^2}{K^2 y^4} + \frac{y^4}{L^4 r^2}  {\cal
  F}_{\tau \sigma}^2  
& = & 0, \nonu \\ 
r'' - \frac{ \left(T_5  \frac{8}{3} \pi^2  L^4 \sin^4
    \theta_k
\right)^2 
L^4 r }{K^2 y^4} + \frac{y^4  {\cal
  F}_{\tau \sigma}^2 }{L^4 r^3}   & = & 0, \nonu \\
y'' + \frac{2 \left( T_5  \frac{8}{3} \pi^2 
    L^4 
\sin^4 \theta_k  \right)^2 L^4 r^2}{K^2 y^5}  - \frac{2y^3  {\cal
  F}_{\tau \sigma}^2}{L^4 r^2}   & = & 0
\label{equationss}
\eea
where we used the constant of equation of motion (\ref{constants}).

The two circular Wilson loops, as in previous section,
 are located at $x=\pm l/2$ on the $AdS_5$ 
boundary $y=0$ and the boundary conditions are given by 
$r(\pm l/2)=R$ and $y(\pm l/2)=0$.
$R$  is the radii of the Wilson loop and $l$ is the distance between them.
By manipulating the equations (\ref{equationss}), one gets 
$(r^2+y^2)'' +2 =0$ which leads to the same solutions in previous section.  
The parametric angle $\theta$ satisfies
\bea
\theta' = \pm \frac{a}{a^2-x^2} \sqrt{\frac{\left( T_5  \frac{8}{3} \pi^2 
    L^6 
\sin^3 \theta_k  \right)^2   \cos^2 \theta}{K^2 a ^2 \sin^4
\theta}-1}
\label{thetaprimes}
\eea
where the upper(lower) sign is the negative(positive) $x$.
For these solutions, the 2-form gauge field has 
\bea
 {\cal F}_{\tau \sigma} = -i \cos \theta_k \left[
\frac{\left(T_5  \frac{8}{3} \pi^2 
    L^4 
\sin^3 \theta_k\right)  
\cos^2 \theta L^4}{K(a^2-x^2) \sin^4 \theta }\right]
\label{F}
\eea
where the quantity inside of the bracket is 
the volume of 2-dimensional induced metric parametrized by
$\tau$ and $\sigma$.





Now it is ready to compute the on-shell action by 
substituting the above solution into the action. 
By changing the integration variable $x$ into $\theta$
and using the relation (\ref{thetaprimes}), one arrives at
\bea
S_{bulk} &=& 2\pi \int_{-\frac{l}{2}}^{\frac{l}{2}}
 dx 
{\cal L}_{bulk} \nonu \\
& = & \frac{N\sqrt{\lambda}}{3 \pi^2} 
\left[ \sin^5 \theta_k -\cos \theta_k \left( \frac{3}{2} \theta_k
-\sin 2\theta_k +\frac{1}{8} \sin 4\theta_k \right) \right] \nonu
\\
&& \times 4\pi \int_{\frac{\epsilon}{R}}^{\theta(x=0) } d \theta 
\frac{ \cot^2 \theta}{\sqrt{\cos^2 \theta - 
\left(\frac{K }{ T_5  \frac{8}{3} \pi^2 
    L^6 
\sin^3 \theta_k}\right)^2 a^2 \sin^4 \theta}}
\label{bulk}
\eea
where we used modified boundary conditions 
$y(\pm l/2)=\epsilon$ and $\theta(\pm l/2)=\tan^{-1} \left( 
\frac{\epsilon}{R}\right)$ as in previous section.
The last line of (\ref{bulk}) is the same expression for the 
area of the connected surface corresponding to 
Nambu-Goto action, found in previous section, 
for a fundamental string on the type IIB background with 
a constant replaced by $\frac{K }{ T_5  \frac{8}{3} \pi^2 
    L^6 
\sin^3 \theta_k}$. 
Then using the same procedure, 
one can integrate explicitly and it leads to
\bea
S_{bulk}&=&  \frac{N\sqrt{\lambda}}{3 \pi^2} 
\left[ \sin^5 \theta_k -\cos \theta_k \left( \frac{3}{2} \theta_k
-\sin 2\theta_k +\frac{1}{8} \sin 4\theta_k \right) \right]
\nonu \\
&& 
\times 4\pi
\left(  
\frac{ (1-\kappa^2) K(\kappa) - 
E(\kappa) }{\sqrt{2\kappa^2-1}}+
\frac{  R}{\epsilon} \right) 
\nonu
\eea
where $\kappa$ is given by
\bea
2\kappa^2 = 1 + \frac{1}{\sqrt{1 +  4 \left( 
\frac{ K  }{ T_5  \frac{8}{3} \pi^2 
    L^6 
\sin^3 \theta_k}\right)^2 a^2  }} 
=1 + \frac{1}{\sqrt{1 +  4 \left( 
\frac{ 3\pi^2 K  }{ N \sqrt{\lambda}
\sin^3 \theta_k}\right)^2 a^2  }}.
\label{kappa}
\eea
Note that when $K=0$, then $\kappa=1$ from (\ref{kappa}) and $E(1)=1$. 
This implies that the half of above on-shell bulk action is equal to the
one of a single circular Wilson loop case \cite{Yama}, as we expected.  

Since the bulk action is divergent, 
this divergent term can be fixed by the boundary terms. 
The only coordinate we have to replace with its momentum is the 
radial coordinate $y$ and it is an integral over the boundary at a cutoff
$y=y_0$. With the explicit expression of (\ref{py}),
we should add the following boundary term 
\bea
S_{body,y} = -\int d \sigma p_y y|
=-2\pi K y y'|=-\frac{2N\sqrt{\lambda}}{3 \pi} \frac{R}{\epsilon}
\sin^3 \theta_k + \cdots
\nonu
\eea
where we used the fact that $\theta(\pm l/2)=\tan^{-1} \left( 
\frac{\epsilon}{R}\right)$ and we ignored the terms that vanish when we take 
$\epsilon \rightarrow 0$ limit. This is the same as the one for the 
single circular Wilson loop case \cite{Yama} after taking $\epsilon 
\rightarrow 0$ limit.

Moreover, there should exist another boundary term. 
The momentum conjugate to the gauge field is given by   
(\ref{pa}) and the corresponding boundary term we should add is
\bea
S_{body,A} &=& -2 \pi \int_{-\frac{l}{2}}^{\frac{l}{2}} 
dx p_A \frac{1}{2\pi \alpha'} 
{\cal F}_{\tau \sigma} \nonu \\
& =&  \frac{N\sqrt{\lambda}}{3 \pi^2}  \cos \theta_k
\left(\frac{3}{2} \theta_k -\frac{3}{4} \sin 2\theta_k \right)
\times
 4\pi
\left[  
\frac{ (1-\kappa^2) K(\kappa) - 
E(\kappa) }{\sqrt{2\kappa^2-1}}+
\frac{  R}{\epsilon} \right]  
\nonu
\eea
where we substituted the expression for 
${\cal F}_{\tau \sigma}$ in (\ref{F}).
The half of above boundary action $S_{body,A}$ is equal to the
one of a single circular Wilson loop case \cite{Yama} when $K=0$, 
as we expected.


Finally, summing up all the contributions, the total action is summarized by
\bea
S_{tot} &= & S_{bulk} + 2S_{body,y} +S_{body,A} 
\nonu \\
&=&
\left( \frac{2N}{3\pi} \sin^3 \theta_k \right) 
2\sqrt{\lambda}
\left[ \frac{(1-\kappa^2) K(\kappa) - 
E(\kappa)}{\sqrt{2\kappa^2-1}} \right] \nonu \\
&=&
4\pi K a \frac{\sqrt{2\kappa^2-1}}{\kappa \sqrt{1-\kappa^2}} 
\left[ (1-\kappa^2) K(\kappa) - 
E(\kappa) \right]
\label{a}
\eea
where we added the boundary term coming from $S_{body,y}$ twice
due to the two circular Wilson loops. 
There is also similar result by looking at how the embedding of
a fundamental string appears in an embedding of D5-brane into
the 10-dimensional background in \cite{Hartnoll}.
For $\kappa =1/\sqrt{2}$ limit (\ref{kappa})(or when $l$ goes to zero),
one gets
\bea
S_{tot}=  - \left( \frac{2N}{3\pi} \sin^3 \theta_k \right) 
2\sqrt{\lambda} \frac{4\pi^3 }{\Gamma^4(
\frac{1}{4})} \frac{R}{l}.
\nonu
\eea
Note that the coefficient term 
of $\frac{2N}{3\pi} \sin^3 \theta_k$ in $S_{tot}$ is exactly the minimal 
surface for anti-parallel lines, each of length $R$ 
and separated by a distance
$l$.

So far we didn't consider Lunin-Maldacena deformation. 
From the experience of
previous section, one can deform the above on-shell action by realizing that
the 10-dimensional metric has an overall factor $\sqrt{H}$ which will give
an extra $H^{3/2}$ factor in the action (\ref{a}) and 
$\kappa$ has ``deformed'' 
expression. The $l$ dependence of $S_{tot}$ resembles the Figure 2 with larger
overall shift of $S_{tot}$ due to the higher power of deformation parameter 
$H$, compared with Figure 2.

\section{Discussion}

For nonconformal theories, the Nambu-Goto action for a fundamental string 
on the type IIB supergravity background for the minimal surface 
can be generalized to
\bea
S = 2\pi \int d \tau \frac{r}{z^2} \sqrt{H L_1 L_2} 
\sqrt{x'^2 + r'^2 + \frac{1}{L_1
L_2^2}
z'^2}, \qquad
L_i = 1+ \ell_i^2 z^2, 
\nonu
\eea
where $\ell_i$ are two parameters specifying the ellipsoidal shape
of D3-brane distribution \cite{KLT}. Of course, the vanishing $\ell_i$ case reduces
to (\ref{nbaction}) for the conformal theory we have discussed so far. 
Then one can construct the Euler-Lagrange equations of motion which is 
a generalization of (\ref{equations}).
Although the analytic solutions are not possible, one expects that
the similar analysis to \cite{KPTM} can be done numerically and by tuning 
the parameters and deformation parameter appropriately, one might see the 
transition to a linearly confining phase.  

In \cite{DF}, 
through a stereographic projection by conformal transformation,
two concentric circles on parallel planes we are considering in this paper  
define a 2-sphere in $\R^4$ with different values for the 5-sphere angle.
It would be interesting to study our analysis here for the nonzero 
difference of 5-sphere angles.
Moreover, 
the generalization of \cite{Zarembo02} in the case of a line or circle 
with periodic motion inside an $\S^2$ is given in \cite{DF} and some periodic
motion inside an $AdS_3 \times \S^3$ subspace is also given. Some of the simple
solution was given already in \cite{TZ}. 
It is natural to ask how the marginal 
deformation plays the role in these examples.

One can consider the $\hat\sigma$ deformations of finite temperature
theories which are described by nonextremal D3-branes. For undeformed case
wih finite temperature case, the analysis was given already in \cite{KPTM}.
The $L_{max}$ and $L_{cri.}$ 
with nonzero temperature are increased as the temperature
increases and the plot of an energy versus the distance implies that the
effect of temperature leads to an enhancement of an area of minimal
surface.

As pointed out in \cite{VP}, it is straightforward to apply the present
description to other types of Ssaki-Einstein spaces for the
5-dimensional internal space. For example, for the $T^{1,1}$ space,
the deformation parameter has similar functional form \cite{VP} except
that the numerical coefficient of $\hat\sigma^2$ in $H$ has different value. 
So we do expect to have similar results.

\vspace{.7cm}

\centerline{\bf Acknowledgments}

We would like to thank Nadav Drukker, DaeKil Park, Justin F. Vazquez-Poritz, 
Satoshi Yamaguchi, and 
Konstantin Zarembo for discussions.
This work was supported by grant No.
R01-2006-000-10965-0 from the Basic Research Program of the Korea
Science \& Engineering Foundation. 
We thank Caltech Particle Theory Group for hospitality where this work 
was undertaken.

\end{document}